\documentclass[pre,10pt, a4paper,superscriptaddress,nofootinbib,twocolumn,longbibliography,showkeys]{revtex4-1}
\usepackage{graphicx}

\usepackage{placeins} % used for /FloatBarrier
\PassOptionsToPackage{hyphens}{url}
\usepackage{hyperref} % For hyperlinks in the PDF
\usepackage{bm}
\usepackage{amsmath} % package for equation
\usepackage[parfill]{parskip} %new paragraph with empty line rather than indent
\usepackage{amssymb,latexsym} %amsthm package extended theorem environments
\usepackage{amsthm}
\usepackage{amsmath}
\usepackage{mathtools}
\usepackage{setspace}
\usepackage{natbib}
\usepackage{url}
\usepackage{breakurl}
\usepackage{float}
\usepackage{xcolor}
\usepackage{cancel}
\usepackage{soul}

\begin{document}
\title{ A Lagrangian model of \textit{Copepod} dynamics: \\ 
clustering by escape jumps in turbulence\\
}

\author{H. Ardeshiri}
\email{hamidreza.ardeshiri@polytech-lille.fr}
\affiliation{Univ. Lille, CNRS, FRE 3723, LML, Laboratoire de M\'ecanique de Lille, F 59000 Lille, France}
\affiliation{Univ. Lille, CNRS, Univ. Littoral Cote d'Opale, UMR 8187, LOG, Laboratoire d'Oc\'eanologie et de G\'eoscience, F 62930 Wimereux, France}
\author{I. Benkeddad}
\author{F. G. Schmitt}
\author{S. Souissi}
\affiliation{Univ. Lille, CNRS, Univ. Littoral Cote d'Opale, UMR 8187, LOG, Laboratoire d'Oc\'eanologie et de G\'eoscience, F 62930 Wimereux, France}
\author{F. Toschi}
\affiliation{Department of Applied Physics and Department of Mathematics and Computer Science, Eindhoven University of Technology, 5600 MB, Eindhoven, The Netherlands}
\affiliation{Istituto per le Applicazioni del Calcolo CNR, Via dei Taurini 19, 00185 Rome, Italy}
\author{E. Calzavarini}
\affiliation{Univ. Lille, CNRS, FRE 3723, LML, Laboratoire de M\'ecanique de Lille, F 59000 Lille, France}
%----------------------------------------------------------------------------------------

\date{\today}

%----------------------------------------------------------------------------------------
%	ABSTRACT
%----------------------------------------------------------------------------------------

\begin{abstract}
Planktonic copepods are small crustaceans that have the ability to swim by quick powerful jumps. Such an aptness is used to escape from high shear regions, which may be caused either by flow perturbations, produced by a large predator (i.e. fish larvae), or by the inherent highly turbulent dynamics of the ocean. Through a combined experimental and numerical study, we investigate the impact of jumping behaviour on the small-scale patchiness of copepods in a turbulent environment. Recorded velocity tracks of copepods displaying escape response jumps in still water are here used to define and tune a Lagrangian Copepod (LC) model. The model is further employed to simulate the behaviour of thousands of copepods in a fully developed hydrodynamic turbulent flow obtained by direct numerical simulation of the Navier-Stokes equations.  First, we show that the LC velocity statistics is in qualitative agreement with available experimental observations of copepods in turbulence. Second, we quantify the clustering of LC, via the fractal dimension $D_2$. We show that $D_2$ can be as low as $\sim 2.3$ and that it critically depends on the shear-rate sensitivity of the proposed LC model, in particular it exhibits a minimum in a narrow range of shear-rate values. We further investigate the effect of jump intensity, jump orientation and geometrical aspect ratio of the copepods on the small-scale spatial distribution. At last, possible ecological implications of the observed clustering on encounter rates and mating success are discussed.
\end{abstract}

\maketitle
%-----------------------------------------
%      INTRODUCTION
%-----------------------------------------
\section{Introduction}

The study of swimming microorganisms and their interaction with fluid flows has attracted enormous attention in the last decade. A line of research  has focused on characterizing individual swimming strategies by means of experiments \cite{Berg-1979, Leonardo-2010, Visser-2006} as well as by theoretical and numerical modelling \cite{Lauga-2016,Teran-2010}.  A second direction of study devoted to the consequences of swimming on population dynamics, e.g., by focusing on encounter rates and other collective behaviours \cite{Hernandez-2005, Baskaran-2009,Lambert-2013,Lushi-2014, Drescher-2011}.  A third direction focused on the mutual interactions of microorganisms with the fluid flow environment, in particular bio-induced flow fluctuations, sometimes dubbed as bacterial turbulence \cite{Hohenegger-2010,Dunkel-2013, Kaiser-2014}, or, vice-versa, on active matter clustering induced by non homogeneous flows or fluid turbulence \cite{Croze-2013, Durham-2013, Pedley-1992, Warnaas-2006, Guasto-2012, Bergstedt-2004, Stocker-2012, Gustavsson-2016, DeLillo-2014}. The present study will focus on this latter aspect, in particular on copepod's dynamics in turbulent flow.\\
Copepods are the most diversified crustaceans in the aquatic environment whose length ranges from $0.1$ mm to few millimetres. They are important to global ecology and to the carbon cycle \cite{Frangoulis-2005} (see also \cite{Satapoomin-1999} \& \cite{Jonasdottir-2015}). Although copepods are not at the top of the food web, they have a major role in the marine ecosystem because they are the secondary producers in the ecological food web linking phytoplankton cells (the primary producers) to fish larvae and even to large mammals such as whales. Copepods also consume the mosquito larvae, acting as control mechanism for malaria \cite{Walter-2015}. They are of great importance in fishery industry. A central issue in breeding fish species, is the external food supply. Most fishes prefer copepods to other zooplankton species (i.e. rotifers) and they grow bigger in shorter time when eating copepods \cite{Theilacker-1984, Souissi-2014}.\\
Living in a fluid environment characterised by body-scale Reynolds number up to 1000, they are subjected to the physics of the flow field both in viscous and inertial regime \cite{Yen-2000}. Copepods typically have a short, cylindrical body with antennas, few pairs of swimming legs and tales. Using their antennas, copepods can sense the disturbance, which is caused either by the presence of predators or by high turbulent regions in the flow. Ki\o rboe et al. \cite{Kiorboe-1999, Kiorboe-1999-2} performed series of experiments, investigating the effect of non uniform flow motion on copepods. In order to find the component of the flow which copepods react the most to, the copepods were put into a time dependent siphon flow (which  ideally generates a pure longitudinal deformation rate), in an oscillating chamber where copepods experience only acceleration, in a couette device producing shear deformation, and finally in a rotating cylinder where acceleration and vorticity are both present. The conclusion of this study was that these small crustaceans react to the flow deformation rate. Ki\o rboe also reported \cite{Kiorboe-book}, that there are two threshold values of the deformation rate: the upper one, around $10\, s^{-1}$, corresponds to either the presence of a predators or to a region where turbulence intensity is high, and the lower one, $1\, s^{-1}$, corresponds to regions in the flow where turbulence intensity is lower or food abundance is not enough for copepods. These tiny crustaceans find themselves at ease in regions in between these two thresholds. To avoid uncomfortable regions, copepods exhibit a rapid escape in the flow which is often dubbed a \textit{jump}. Buskey et al. \cite{Buskey-2002, Buskey-2003} showed that copepod's velocity can reach the rate of 500 body length per second (0.5 $m/s$) while jumping. The mechanical energy produced during their escape is reported to be very high ($8\times 10^{-5}\, J/s$) \cite{Lenz-1999}, which makes copepods, relative to their size, among the fastest and the strongest animals in the world.\\
Buskey \cite{Buskey-2002, Buskey-2003} also reported that males and females respond differently to hydrodynamic stimulus in terms of response latency, jump speed, number of thrusts, distance jumped and many other parameters. According to their investigations, copepods jump in an unpredictable direction, but rarely in the backward direction of their motion. Other studies have considered the mating behaviour of copepods \cite{Lee-2011} and the effect of salinity on copepod's dynamics and copepod's encounter rate \cite{Souissi-2010, Schmitt-2008}. Copepods are also sensitive to light stimuli, being attracted by natural light sources \cite{Fields-2012}.\\
In the last two decades many studies have been conducted to quantify the dynamics of copepods. Most of them focused on their behaviour in still water \cite{Lee-2011, Souissi-2010, Schmitt-2008}, while less studies have studied the dynamics in their natural living environment because of the difficulties of such experimental investigations. Few works have been devoted to the dynamics of copepods in turbulent flows \cite{Moison-2009, Waggett-2007, Yen-2008, Michalec-2015, Michalec-2015-2}. However, the densities of copepods used in these studies are often lower than the maximum densities that can be encountered in the field.\\ 
The numerical simulation can provide a tool that integrate our current knowledge on copepod dynamics and use high number of individuals. The objective of the present study is to simulate copepods numerically in turbulence to characterise their dynamics induced by a behaviour model. To achieve this goal, our strategy is two-fold: on one hand, new experimental measurements and  observations available in the literature \cite{Jiang-2004, Jiang-2008, Jiang-2011, Kiorboe-2010, Yen-1992, Lenz-2004, Duren-2003}, along with the aforementioned copepods properties, should be considered in details in order to introduce a realistic and physical model. On the other hand, fundamental knowledge on simulation of particles in turbulent flows, available in numerical and experimental studies on particles in turbulence \cite{Toschi-2009, Voth-2009, Parsa-2012, Chevillard-2013,  Zhan-2014}, is needed to couple the physics and biology in the numerical model.\\
The paper is organised as follows: section \ref{Sec:Method} describes the experimental framework used to stimulate copepods. We then analyse copepod's trajectories to introduce a model equation describing copepods behaviour. Furthermore similarity analysis is performed to tune the LC model and its numerical implementation is explained at the end of this section. Section \ref{Sec:Result} details the single point statistics, fractal dimension and orientation dynamics of copepods. The paper ends with conclusion and outlook on future works.
%-----------------------------------------
%      METHOD
%-----------------------------------------

\section{Methods}\label{Sec:Method}

\subsection{Experimental jump data analysis}
\label{Experimental jump data analysis}
\begin{figure}[!htb]
\begin{center}
  \includegraphics[scale = 0.62]{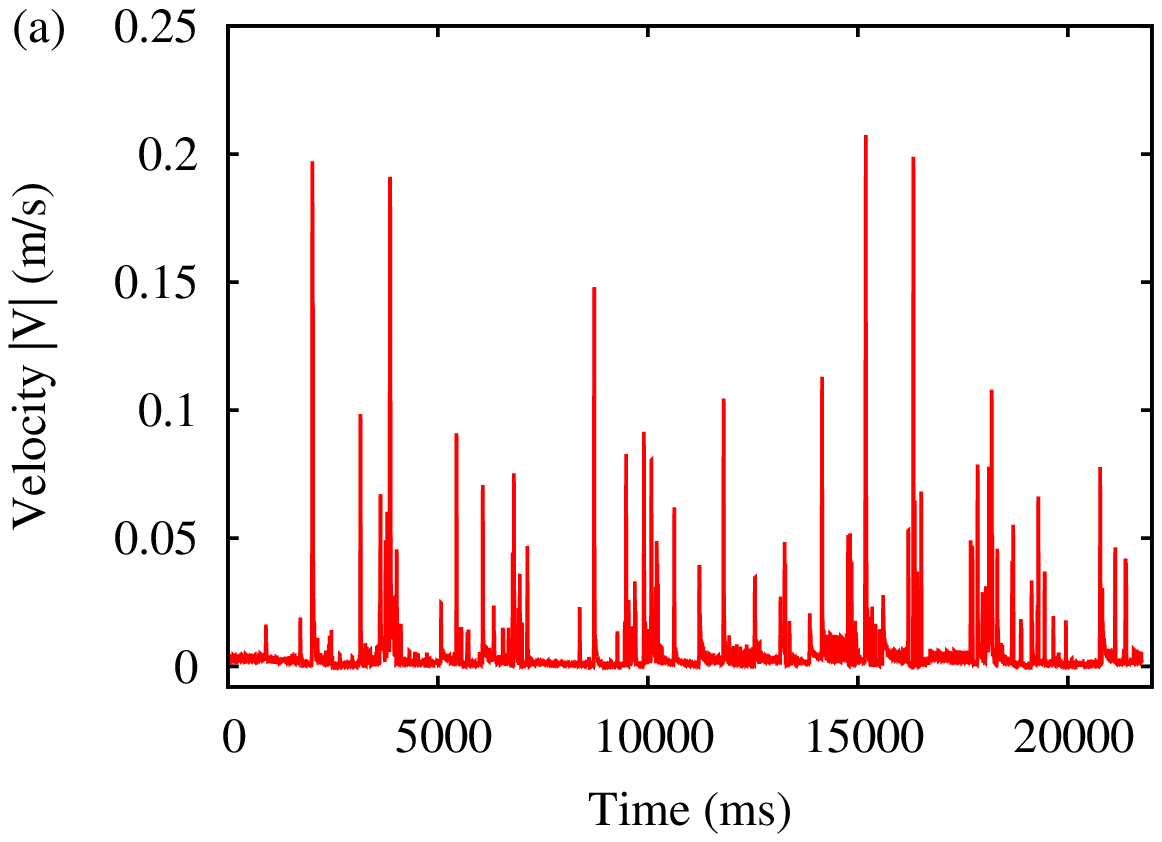}
  \includegraphics[scale = 0.62]{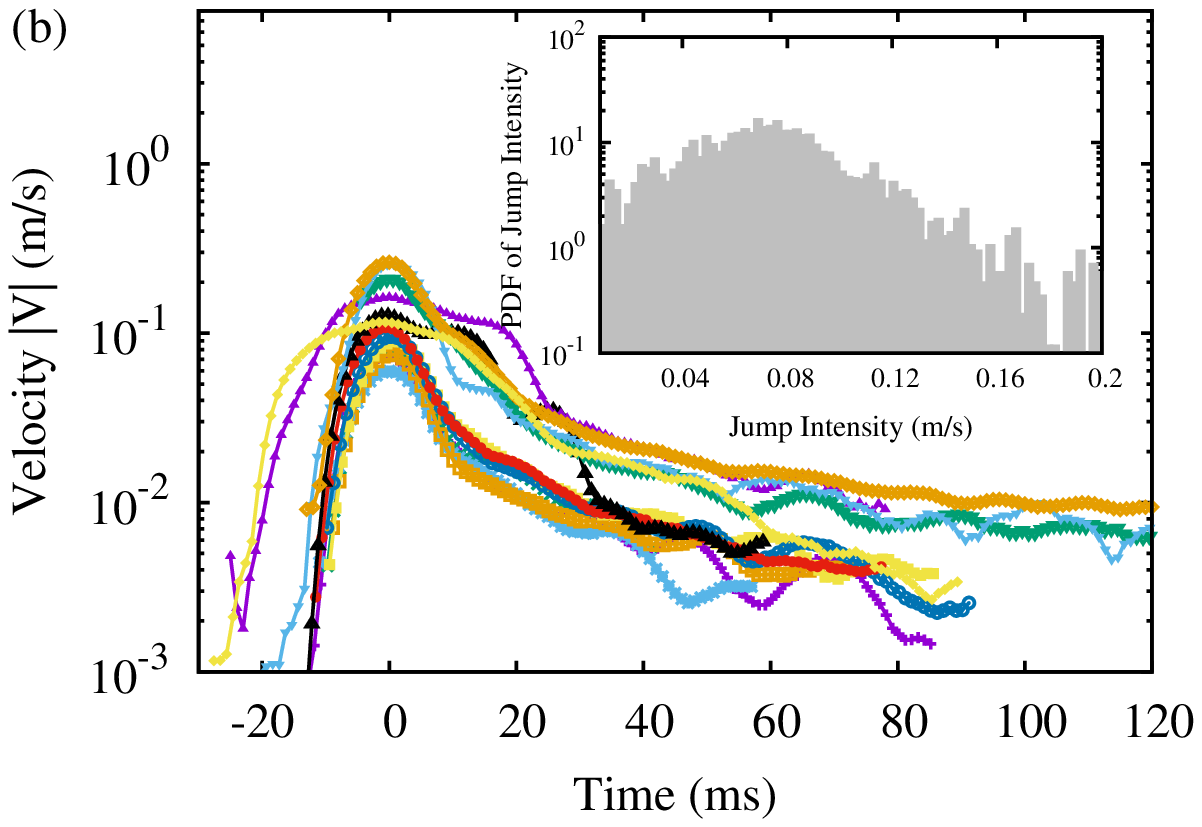}
\caption{(a) The copepod velocity relative to temporal sequence with multiple jumps occurred in response to stimulus. (b) Several jumps superposed by a shift, taking as reference time that are associated with their peak position. Almost all of the jumps decay exponentially. (Inset) The probability density function (PDF) of the jump intensity.}
\label{time-velo-raw}
\end{center}
\end{figure}

\begin{figure}[!htb]
\begin{center}
  \includegraphics[scale = 0.62]{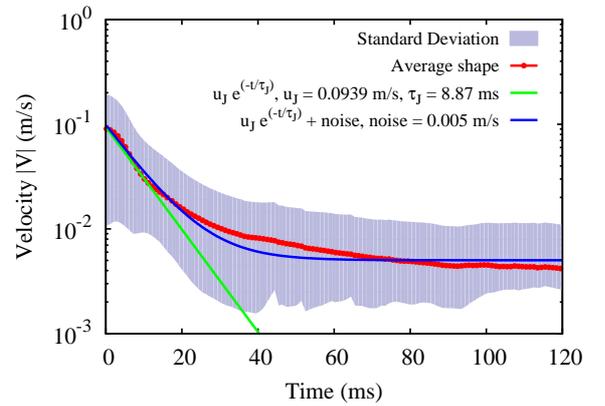}
\caption{Average shape of copepod velocity (over about 730 jumps):  mean value (red line) and standard deviation (shaded area). Note that an error along the horizontal direction due to the uncertainty in the identification of jump peaks may be present but has been omitted here. Green line: Fitted exponential function $u_J e^{-t/\tau_J}$ where $u_J$ is the jump intensity and $\tau_J$ is the decaying time of the jump.
Blue line: Same fit with the addition of a noise velocity offset.}
\label{time-velo-uni}
\end{center}
\end{figure}

We begin presenting an analysis of a new experimental trajectory data set of the estuarine copepod, \textit{Eurytemora affinis}, recorded at LOG Laboratory between May and June 2015. Copepods originated from the Seine river estuary (France) are maintained in the laboratory under optimal conditions for several generations.  
The experimental set up is a shallow-depth aquarium, $63\times53\times6 \,mm^3$ in length, height and depth respectively, with two light sources on the lateral side ($53\times6 \,mm^2$). 
The water is kept still and at temperature of $(18\pm 1) \,^{\circ}\mathrm{C}$ and salinity of $15\ psu$. Copepods were introduced one at a time in the aquarium and their dynamics filmed. A total of 14 individuals were analysed (7 males and 7 females).
A copepod in the aquarium is lead to jump preferentially along the horizontal direction by switching on just one of the light sources.
The copepod dynamics in a vertical plane is recorded by a high speed camera (1000 frames/second) and the single trajectory  is extracted by means of a particle tracking velocimetry software (TEMA Motion by Image Systems).
In such a way hundreds of trajectories are recorded, each with an average time length of $19\ s$. A typical copepod velocity signal as a function of time is shown in figure \ref{time-velo-raw}(a). We see extremely abrupt spikes (jumps) alternating to calm, nearly immobile, phases.\\
In order to see if the velocity signal of the jump events share some common features, we zoom in on the signal and superpose several jumps by a shift taking as reference their peak position. In figure \ref{time-velo-raw}(b) we can appreciate that almost all of the jumps, after a steep rise, display a similar decay.  We associate such a decay to a purely hydrodynamical effect. It can be interpreted as a drag-induced decay of an instantaneous acceleration. 
The inset of panel in figure \ref{time-velo-raw}, shows that the probability density function (PDF) of the jump intensity, has a maximum value around $\sim 0.07 \, m/s$. Note that spikes are identified based on a threshold on the time-averaged velocity of the copepods in each copepod's trajectory.
We then average the dataset of jumps in order to obtain an averaged shape of jump. This is shown in figure \ref{time-velo-uni}, from which we can deduce the average jump velocity amplitude $u_J = 0.0939\, m/s$ and the mean decaying time $\tau_J = 8.87\, ms$. We also see that for long time the velocity reaches a very low plateau at $5 \times 10^{-3}\, m/s \sim 1/20\ u_{J} $, which we are tempted to associate to a weak random wandering behaviour of the copepod.\\
The distribution of jumps in time in the experimental dataset seems to deviate from an exponential distribution suggesting the existence of a memory effect. This may however be dependent on the type of stimulus (the light source) which is continuous in time, very different from the one due to the presence of a variable flow shear-rate.  This aspect will therefore not be taken into account in  the model presented in the next section. We plan to investigate inter-jump statistics more carefully in the future, when experiments with mechanically induced stimulus may be available.

%\FloatBarrier

\subsection{Model equation for copepods dynamics}
In this section we introduce a simple model system of copepod's dynamics. This representation is based on the idea that the copepod's trajectories in a fluid can be mimicked by properly defined active particles. Similar models have been successfully employed for the description of the behaviour of phytoplankton, such as chlamydomonas \cite{Stocker-2009, Kessler-1985, Pedley-1992} both in laminar and, more recently, in turbulent flows \cite{DeLillo-2014, Durham-2013, Croze-2013}. Copepods, and zooplankton in general, display higher complexity compared e.g. to algae species because of their higher motility. 
The model relies both on biological and hydrodynamical assumptions. First, we assume that copepods respond always in the same way to external flow disturbances. Their jump reaction is embedded in their neural system. Furthermore, the stimulus triggering the jump is highly stylised, we only take into account a mechanical signal with a single-threshold, to be specified later on,  and ignore any other activity induced by light, food, or chemistry (e.g. pheromones).
On the mechanical side, we assume that copepods are small enough that their centre of mass can be considered a perfect fluid tracer in a flow, except for the time when a jump event takes place. In hydrodynamic terms this means that copepods are assumed to be rigid, homogeneous, neutrally buoyant particles with a size which is of the order of the dissipative scale of the flow. Gravity force has no role in producing acceleration or torque. Only the drag force effect is taken into account during the jumps.
Finally, copepods are coupled to the fluid in a one-way fashion, they react and are carried by it, but they do not modify the surrounding flow, copepods-copepods interactions are also neglected.
Adding all together the above hypothesis the LC equation of motion is as follows:
\begin{equation}
\dot{\bm{\mathrm{x}}}(t)= \bm{\mathrm{u}}(\bm{\mathrm{x}}(t),t) + \bm{J}(t,t_i,t_e,\dot{\gamma},\bm{\mathrm{p}})
\label{xdot}
\end{equation}\\
where $\bm{\mathrm{u}}(\bm{\mathrm{x}}(t),t)$ is the velocity of the carrying fluid at time $t$ and position ${\bm{\mathrm{x}}(t)}$ and where $\bm{J}$ is an added velocity term that describes the active behaviour (jump) of the copepod. $\bm{J}(t,t_i,t_e,\dot{\gamma},\bm{\mathrm{p}})$ is a function of time $t$, it depends also on an initial and a final time $t_i$ and $t_e$, on flow shear rate value $\dot{\gamma}$  and on orientation vector $\bm{\mathrm{p}}$.  
If copepods are taken to be spherical in shape, their orientation dynamics is given by:
\begin{equation}
\dot{\bm{\mathrm{p}}}(t) = \bm{\Omega}  \cdot  \bm{\mathrm{p}}(t)
\label{eq-pdot}
\end{equation}\\
where $ \bm{\Omega}$ is the fluid rotation rate antisymmetric tensor, defined as $\Omega_{ij} = 1/2 (\partial_i u_j - \partial_j u_i)$. A more general form of the equation (\ref{eq-pdot}), valid for axisymmetric ellipsoidal particles, is as follows:

\begin{equation}
\dot{\bm{\mathrm{p}}}(t) = \left( \bm{\Omega}  + \tfrac{\alpha^2-1}{\alpha^2+1}
\left(\mathcal{S} - \bm{\mathrm{p}^T}(t) \cdot \mathcal{S} \cdot \bm{\mathrm{p}}(t)  \right)\right)  \cdot  \bm{\mathrm{p}}(t)
\label{eq-pdot-jeffery}
\end{equation}

where $\alpha \equiv l/d$ is the aspect ratio of the ellipsoids given by the ratio of length ($l$) to diameter ($d$), which is typically around 3 for \textit{Eurytemora affinis}. The above equation was first proposed by Jeffery, and its full derivation is detailed in \cite{Jeffery-1992}. Its phenomenology in turbulent flows has been investigated more recently in \cite{Parsa-2012}.
Notice that here we designate by $\mathcal{S}$ the fluid deformation rate symmetric tensor as $S_{ij} = 1/2 (\partial_i u_j + \partial_j u_i)$ and the shear rate is then defined as $\dot{\gamma} = \sqrt{2\,\mathcal{S}:\mathcal{S}}$. We note that the fact that the jump term is assumed to depend on $\dot{\gamma}$ represents a generalization to the 3D geometry of Ki\o rboe's empirical findings \cite{Kiorboe-book}.
For the jump term we propose the following functional form:
%\begin{widetext}
\begin{equation}
% space in equation   \,    \;   \:   \quad
%\bm{J}(t,t_i,t_e,\dot{\gamma} ,\bm{\mathrm{p}}) = H[\dot{\gamma}(t_i) - \dot{\gamma}_T]\, H[t_e - t]\, u_{J}\, e^{-\frac{(t-t_i)}{\tau_{J}}}\, \bm{\mathrm{p}}(t_i)
\bm{J}(t,t_i,t_e,\dot{\gamma} ,\bm{\mathrm{p}}) = H[\dot{\gamma}(t_i) - \dot{\gamma}_T]\, H[t_e - t] u_{J}\, e^{\frac{t_i-t}{\tau_{J}}}\, \bm{\mathrm{p}}(t_i)
\end{equation}
%\end{widetext}
where $H[x]$ denotes the Heaviside step function, $\dot{\gamma}_T$ is a threshold value of the shear rate, $u_{J}$ and $\tau_{J}$ are two characteristic parameters characterising the jump shape, its velocity amplitude ($u_J$) and duration $\tau_J$ respectively. The first $H$ step function models the fact that a jump can begin only when the shear rate is above the given threshold value, while the second step function accounts for the fact that the jump time span is finite. The initial and final time of a jump are defined as:
\begin{eqnarray}
t_i = t \quad \textrm{if} \quad (\dot{\gamma}(t) > \dot{\gamma}_T) \cap (t > t_e) \\
t_e = t_i  +  c \, \tau_{J} = t_i +  \textrm{log}(10^2) \, \tau_{J}
\end{eqnarray}\\
In other words we assume that a jump can not begin if a previous jump has not finished ($t > t_e$) and that a jump terminates when its amplitude has decreased to a negligible level, here taken as one percent of the initial amplitude, i.e.\ $|\bm{J}(t_e)| = 10^{-2} |\bm{J}(t_i)|$.

%-----------------------------------------
%      METHOD - subsection
%-----------------------------------------
\subsection{Model tuning for turbulent flows}
We now take into account the presence of  the oceanic flow environment surrounding the copepods.  The properties of oceanic turbulence relevant for our work have been studied, among others by  MacKenzie et al. \cite{MacKenzie-1993} and Jimenez \cite{Jimenez-1997}. In these surveys it was observed that the mean value of the turbulent kinetic energy dissipation rate, $\epsilon = 2\nu \mathcal{S}:\mathcal{S}$, varies from about $10^{-8}\,m^{2}s^{-3}$ in open ocean to $10^{-4}\,m^{2}s^{-3}$ in coastal zones (although it is also sensitive  to the wind speed conditions and on the depth).  The value of $\epsilon$ along with the kinematic viscosity of sea water, $\nu$, allow to estimate the Kolmogorov scales of ocean turbulence: The dissipative length ${\eta} = (\nu^3/\epsilon)^{1/4}$,  time $\tau_\eta = (\nu \epsilon)^{1/2}$ and velocity $u_{\eta} = (\nu \epsilon)^{1/4}$. The order of magnitude estimate as from Ref. \cite{Jimenez-1997}  for these quantities are reported in table \ref{ocean-water}.  According to the same authors the typical Taylor-scale Reynolds number $Re_\lambda$ in the ocean can reach values up to $\mathcal{O}(10^2)$.\\
Given that the typical size of copepods is of the order of millimetres, it is clear that the relevant flow scales for their dynamics are close to the Kolmogorov scale or below in turbulence \cite{Yen-2000}.
When the LC model is recast in a dimensionless form in terms of these scales we get three dimensionless groups of parameters:  $\tau_J/\tau_\eta$, $u_J/u_\eta$ and $\tau_\eta \dot{\gamma}_T$.  These parameters, together with the flow  $Re_\lambda$  fully specify the working conditions (or tuning) of the copepods-in-turbulence model. \\
In this study  we take as reference for the energy dissipation rate the value $\epsilon = 10^{-6}\,m^{2}s^{-3}$, and by taking into account the dimensional values estimated for the copepods jump intensity $u_J$ and jump decaying time $\tau_J$ , the ratios $u_J/u_\eta = 93.9$ and $\tau_J/\tau_\eta = 0.00887$ can be deduced from the similarity analysis. 
This tells us that in ordinary turbulence conditions the copepods possess an almost instantaneous reaction, since their response time is about one hundredth of the smallest scale of turbulence. On the opposite the velocity reached during a jump is of a magnitude that  is comparable if not higher to the one of turbulent velocity fluctuations.
Finally, we note that we do not have any experimental guess for the magnitude of $\dot{\gamma}_T$, therefore the value  $\tau_\eta \dot{\gamma}_T$ is a free parameter of our model.

%TABLE
\begin{table}[!htb]
\begin{center}
%\begin{tabular}{| c | c | c | c | c | c |}  % number of centered column)
\begin{tabular}{| c | c | c | c | c |}
\hline %\hline                       % 2 horizontal line
%\nu   & \epsilon & \eta & \tau_{\eta} & u_{\eta} & Re_{\lambda}\\ [0.5ex] % heading
Parameter & Unit &\multicolumn{2}{c|}{Range} & This study\\
\hline
%10^{-6} (m^{2}s^{-1}) & 10^{-6} (m^{2}s^{-3}) & 1 (mm) & 1 (s) &  1 (mm s^{-1}) & O(10^2)\\
$\nu$ & $m^{2}s^{-1}$ & \multicolumn{2}{c|}{$\sim 10^{-6}$} &$10^{-6}$\\
\hline
$\epsilon$ & $m^{2}s^{-3}$ & $10^{-8}$ & $10^{-4}$ &$10^{-6}$\\
\hline
$\eta$ & $m$ & $3\times10^{-3}$ & $3\times10^{-4}$ &$10^{-3}$\\
\hline 
$ \tau_{\eta}$ & $s$ & $10$ & $0.1$ & $1$\\
\hline
$u_{\eta}$ & $m s^{-1}$ & $3\times10^{-4}$ & $3\times10^{-3}$ &$10^{-3}$\\
\hline
$Re_{\lambda}$ & -- & \multicolumn{2}{c|}{$\mathcal{O}(10^{2})$} &  80 \\
\hline
\end{tabular}
\end{center}
\caption{Reference properties of the ocean turbulent flow as from \cite{Jimenez-1997}. $\epsilon$ is the mean turbulent energy dissipation rate, $\eta$, $\tau_{\eta}$ and $u_{\eta}$ are the turbulence space, time and velocity dissipative scales. $Re_{\lambda}$ is the Taylor-scale based Reynolds number. Their approximate range of variability  is given together with the reference values chosen for the similarity analysis in the present study.}
\label{ocean-water}
\end{table}%
%\FloatBarrier

%-----------------------------------------
%      METHOD - subsection
%-----------------------------------------
\subsection{Numerical implementation of the LC model and of the turbulent flow simulation}
The copepods-in-turbulence model system is conveniently implemented via an Eulerian-Lagrangian approach, meaning that the trajectory $\bm{\mathrm{x}}(t)$ of each individual copepod is computed by means of  Lagrangian tracking method applied to eq.  (\ref{xdot}) \cite{Squires-1990, Elghobashi-1993}, while the fluid flow is obtained by solving the field equations of incompressible fluid-dynamics, i.e. Navier-Stokes equations, in turbulent conditions.
All the particles are advanced in time using Adams-Bashforth method with a time step equal to $\delta t = 1.4 \times 10^{-3} \tau_{\eta}$, the same time step as for the integration of the Navier-Stokes equations. Such a choice of time step shall also satisfy the constraint $\delta t \ll \tau_{J}$.

A Direct Numerical Simulation (DNS) approach was used to solve the Navier-Stokes equations for homogeneous isotropic turbulence by means of a pseudo-spectral method:
\begin{equation}
\begin{split}
%\partial_t{\bm{\mathrm{u}}(\bm{\mathrm{x}}(t),t)} + \bm{\mathrm{u}}(\bm{\mathrm{x}}(t),t) \cdot \nabla{ \bm{\mathrm{u}}(\bm{\mathrm{x}}(t),t)} = \\
% -\nabla{p} + \nu  \Delta{\bm{\mathrm{u}}(\bm{\mathrm{x}}(t),t)} + f
 \partial_t{\bm{\mathrm{u}}} + \bm{\mathrm{u}} \cdot \nabla{ \bm{\mathrm{u}}} =  -\nabla{p}/\rho + \nu  \Delta{\bm{\mathrm{u}}} + \bm{f}
\end{split}
\end{equation}
where $\bm{\mathrm{u}}(\bm{\mathrm{x}}(t),t)$ is the incompressible ($ \nabla \cdot \bm{\mathrm{u}}= 0$) fluid velocity field, $p$ is the pressure, $\nu$ is the kinematic viscosity and $\rho$ is the fluid density. The $\bm{f}$ is the forcing which is applied on large scales to sustain the statistically stationary turbulence. The solution domain is a cube of length $L=2\pi$ with $N^3 = 128^3$ grid points, subject to periodic boundary condition. Aliasing error is controlled by omitting the wavenumber larger than $k=2/3 \times (2\pi N/L)$, to reach the Taylor Reynolds number of the flow $Re_{\lambda} =\sqrt{15}u_{rms}^2/(\nu \epsilon)^{1/2} \approx 80$ where $u_{rms}$ is the single component root mean square velocity fluctuation. $k_{max}\eta > 1.4 $, in which $k_{max} = N/3$ and $\eta$ is the Kolmogorov length scale, assures that small scales structures are well resolved.
%----------------------------------------
%      RESULTS
%-----------------------------------------
\section{Results and Discussion}\label{Sec:Result}
As mentioned above, the LC model is characterised by three control parameters: the jump intensity $u_J$, the decaying time of the jump $\tau_J$ and the shear rate threshold value $\dot{\gamma}_T$, which are conveniently presented in dimensionless form in terms of turbulence dissipative scale units.  Since the LC model is just one-way coupled to the fluid, in the numerics we can perform simultaneous simulations of several families of copepods in the same turbulent flow, where each family is characterized by the triplet $\left[ u_J/u_{\eta} , \tau_J/\tau_\eta , \dot{\gamma}_T \tau_\eta \right]$.\\
In agreement with the experimental observation we always keep fixed the decaying time of the jump to the value $\tau_J/\tau_\eta = 10^{-2}$, while the other parameters are varied 
independently  in the ranges $ u_J/u_\eta \in \left[1,400 \right]$  and $ \tau_\eta \dot{\gamma}_T \in \left[0, 4\right]$.
Note that if $\dot{\gamma}_T = 0$, according to the model, all the particles will jump in a synchronous way. In order to avoid such an unphysical feature, the time $t_e$ for each particle is initialised by a random variable with homogeneous distribution in the interval $[0, \textrm{log}(10^2) \, \tau_{J}]$.
We perform a series of simulations with multiple families, with about  $2.56\times 10^5$ particles per family\footnote{In physical dimension this corresponds to a number density of $\mathcal{O}(1)$  LC particles per $cm^3$, a density comparable to the one found for real copepods estuarine water.}. The simulation was started and particles were let displace for about 2 eddy turnover times, after that during the following $\sim$2 eddy turnover times  about 10 instantaneous distributions of LC particles were saved for analysis.
Copepods are modelled as solid sphere particles, and orientation vector affected by fluid rotation rate (eq. \ref{eq-pdot}), unless otherwise noted. 
For comparison a set of passive fluid tracers are also included in all our simulations.

\subsection{Single Point Statistics}
\begin{figure}[!htb]
\begin{center}
%%%   trim=top  left bottom right 
%  \includegraphics[clip, trim=2.5cm 3.5cm 3cm 3cm, scale = 0.25, angle=-90]{fig1-pdf-ux-abs-case1.eps}  
 % \includegraphics[clip, trim=2.5cm 3.5cm 3cm 3cm,scale = 0.25, angle=-90]{fig1-pdf-ux-abs-case2.eps}  
   \includegraphics[scale = 0.33, angle=-90]{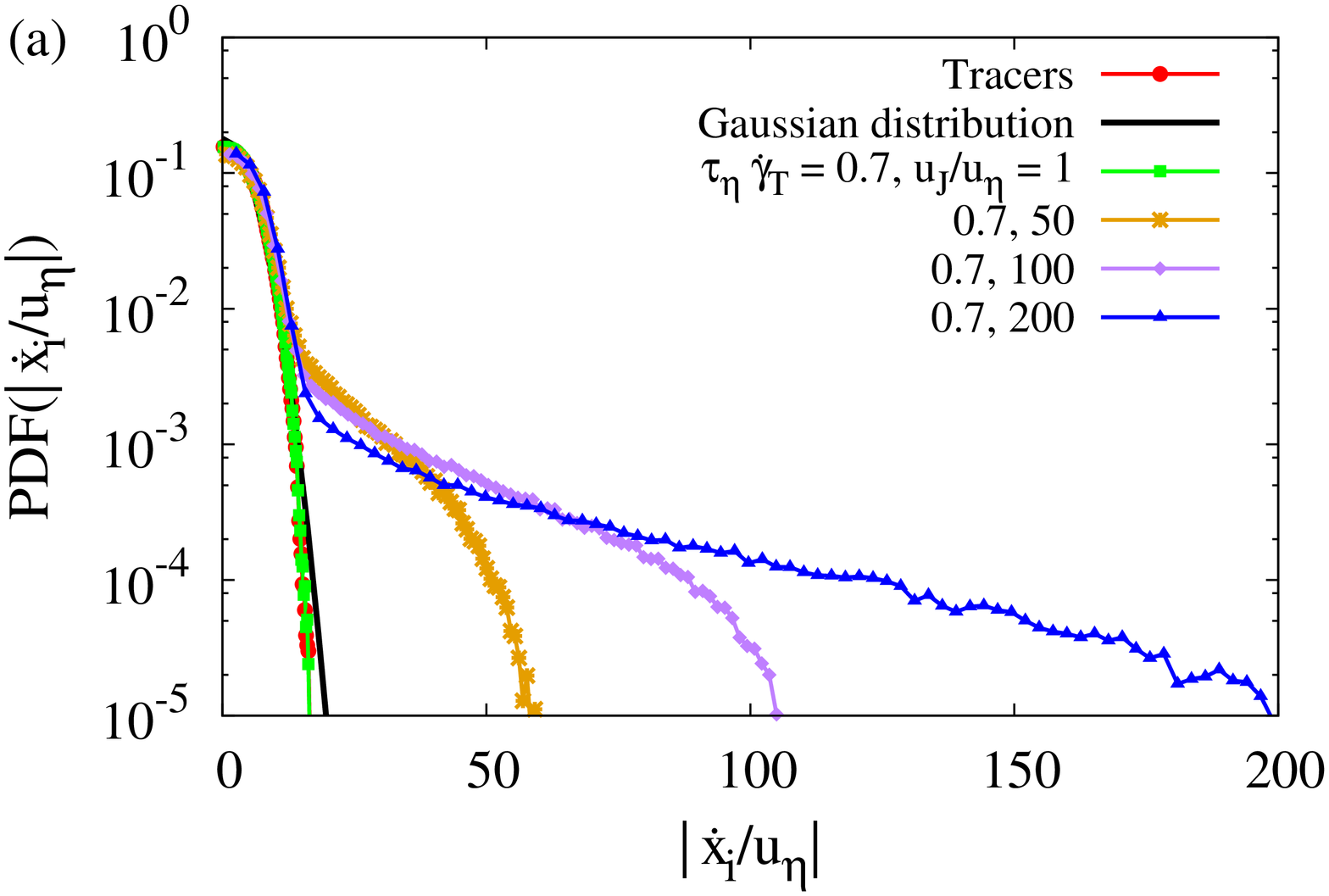}  
   \includegraphics[scale = 0.33, angle=-90]{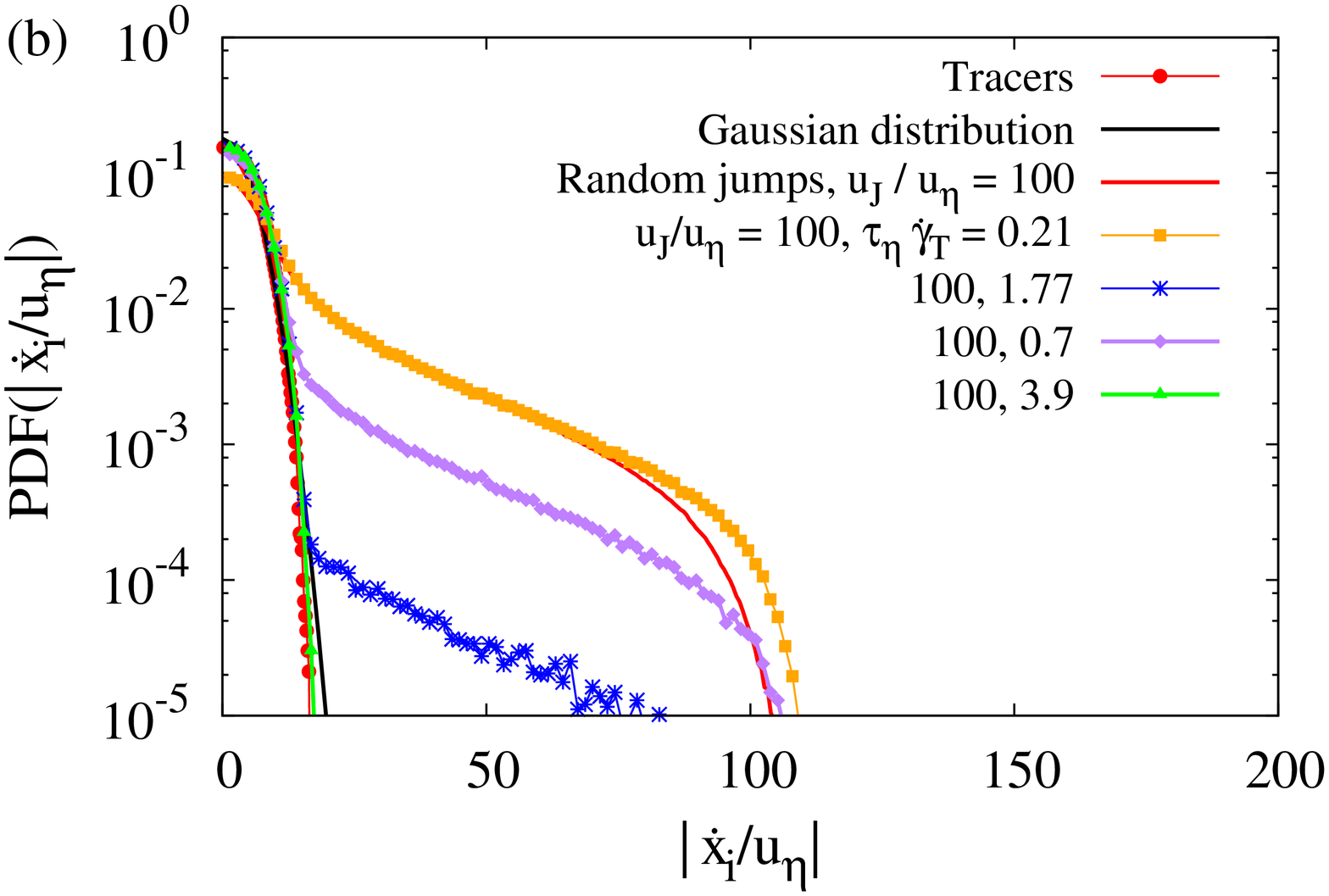}  
\caption{Probability density function of absolute value of single component velocity $|\dot{x}_i/u_\eta|$ for the copepods. (a) at constant threshold value $\tau_\eta \dot{\gamma}_T = 0.7$ and different jump intensities. Gaussian distribution is a statistic distribution here with the measured root mean square velocity of the Eulerian field as the standard deviation. (b) at constant jump intensity $u_J/u_\eta = 100$ for different shear rate threshold values. Random jumps correspond to the expected velocity distribution when randomly oriented jumps occur uniformly in time on top of the turbulent velocity field.}
\label{PDF-abs-velo}
\end{center}
\end{figure}

\begin{figure}[!htb]
\begin{center}
%%%   trim=top  left bottom right 
%  \includegraphics[clip, trim=2.5cm 3.5cm 3cm 3cm, scale = 0.25, angle=-90]{fig1-pdf-ux-abs-case1.eps}  
   \includegraphics[scale = 0.33, angle=-90]{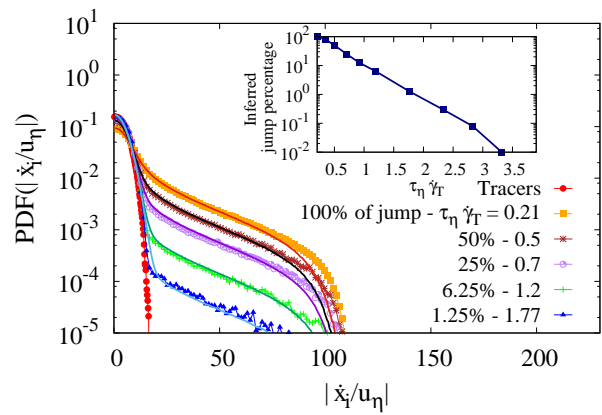}   
\caption{Probability density function of absolute value of single component velocity $|\dot{x}_i/u_\eta|$ for the copepods at constant jump intensity $u_J/u_\eta = 100$ for different shear rate threshold values. Fitted PDF curves correspond to the percentage of jump of copepods. (Inset) Deduced percentage of jump as a function of the shear rate threshold value $\tau_\eta \dot{\gamma}_T$.}
\label{PDF-abs-velo-percentage}
\end{center}
\end{figure}

In order to see how the LC dynamics in turbulence differ from that of a fluid tracer, we first address the velocity single point statistics.
The PDF of the absolute value of single component velocity for the copepods, i.e. $\left| \dot{x}_i \right|$, is shown in figure \ref{PDF-abs-velo}. Tracers, the particles which move along the streamlines, agree with a Gaussian distribution, while for copepods a slower decaying tail is found.
This deviation becomes more pronounced at increasing the jump intensity for a given threshold value of the shear rate, as shown in figure \ref{PDF-abs-velo}(a).
It also appears that low jump intensities $u_J < 10 \,u_{\eta}$ are not strong enough to make effective changes on the copepods PDF. On the other hand, increasing the threshold value of the shear rate leads to fewer jumps, therefore in this case copepods behave almost like tracers. Their deviation in velocity distribution from the Gaussian, indeed increases by decreasing the shear rate threshold value as can be seen in figure \ref{PDF-abs-velo}(b).\\ 
The general trend of the observed deviation from Gaussianity can be predicted by means of the following probabilistic model. We suppose that the instantaneous single cartesian component velocity of LC particles can be approximated by the sum of three statistical independent random variables. The first variable accounts for the turbulent velocity field contribution, therefore it is a Gaussian with zero mean and same standard deviation as the one measured in the DNS. The second and the third variable mimic respectively the jump direction and its intensity: we assume that the orientation is random uniform in the solid angle and that the jumps happen uniformly in time. One can obtain the resulting PDF for the LC particle velocity from the convolution of the three elementary PDFs associated to the three described random variables.
The resulting density distribution function when compared to the LC measurements at low threshold value $\tau_\eta \dot{\gamma}_T = 0.21$ (i.e. when copepods jump very frequently), shows an overall qualitative agreement with a slight deviation in the tails (see Fig.  \ref{PDF-abs-velo}(b)).\\
Such a discrepancy comes from the fact that in reality the jump directions develop some correlations with the underlying flow, via Eq. \ref{eq-pdot}, while the probabilistic model neglects it. One can make use of the approximate probabilistic model to estimate the average fraction of particle performing jumps as a function of the shear-rate threshold value. This is done by introducing an adjustable parameter accounting for the probability that a given particle is actually jumping, and by fitting the model to the PDF curves.
Figure \ref{PDF-abs-velo-percentage} shows the fitted predictions obtained with such a procedure (which confirm the validity of the probabilistic model), while the inset of  the same figure displays the inferred jump percentage as a function of the shear rate threshold value. We observe an exponential decrease as $ \dot{\gamma}_T$ is raised. For the value $\tau_\eta \dot{\gamma}_T = 0.5$, the jumping particle fraction is around 50\%.\\ 
We finally observe that the shape of the PDF displayed by the LC model, is also in qualitative agreement with a recently published experimental study \cite{Michalec-2015}, despite the fact that the experiment has been performed in low Reynolds number conditions (up to $Re_{\lambda} \simeq 30$).
What has not been reported yet in experimental studies is a quantification of the three-dimensional spatial distribution of copepods in turbulence. We do this in the next section by means of a fractal dimension characterization.
%\FloatBarrier

%\FloatBarrier

\subsection{Correlation Dimension Analysis}
\begin{figure}[!htb]
\begin{center}
%%%   trim=top left bottom right 
  \includegraphics[clip, trim=2cm 5.8cm 1.5cm 3cm, scale = 0.45, angle=-90]{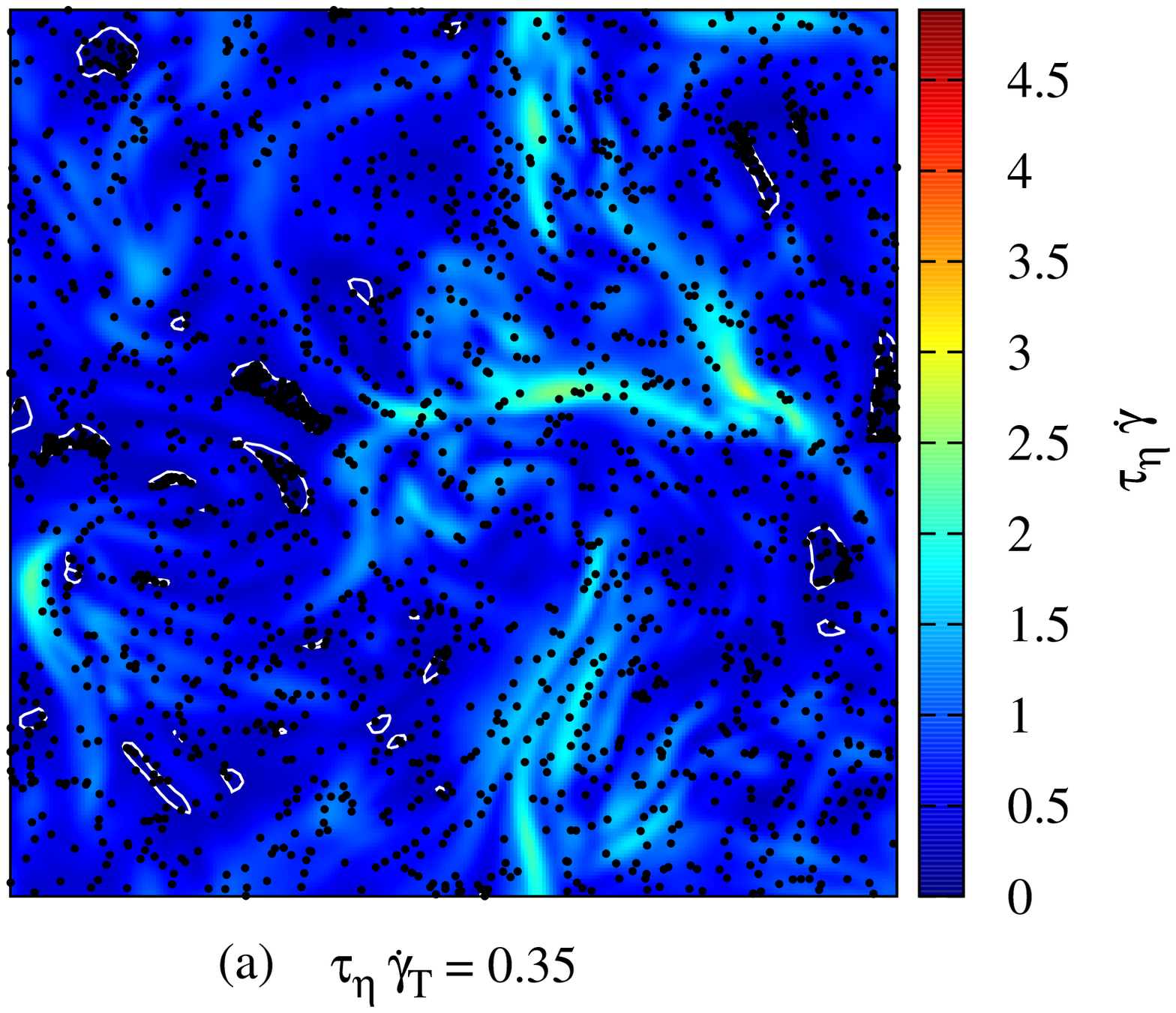}  
  \includegraphics[clip, trim=2cm 5.8cm 1.5cm 3cm, scale = 0.45, angle=-90]{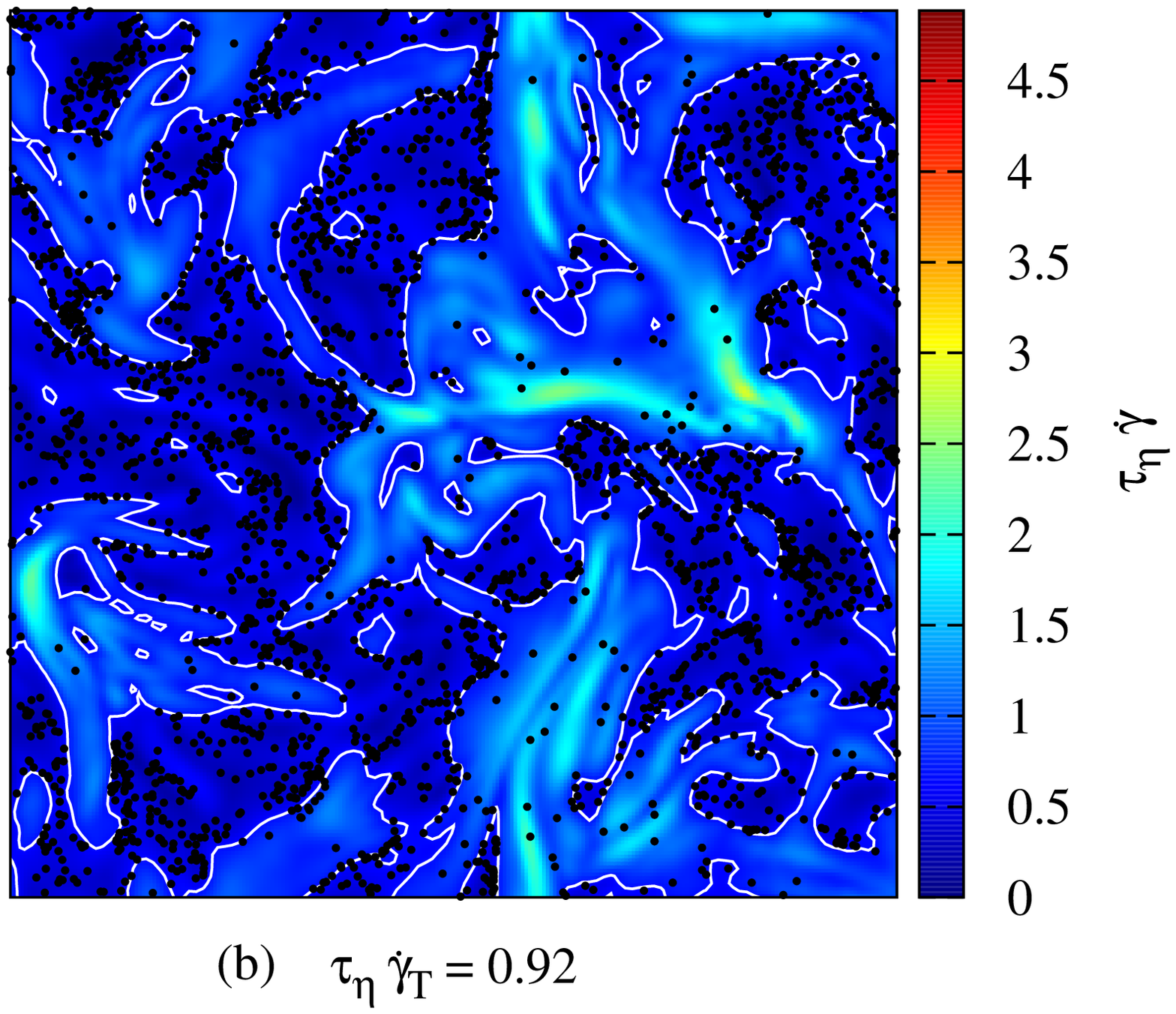}  
  \includegraphics[clip, trim=2cm 5.8cm 1.5cm 3cm, scale = 0.45, angle=-90]{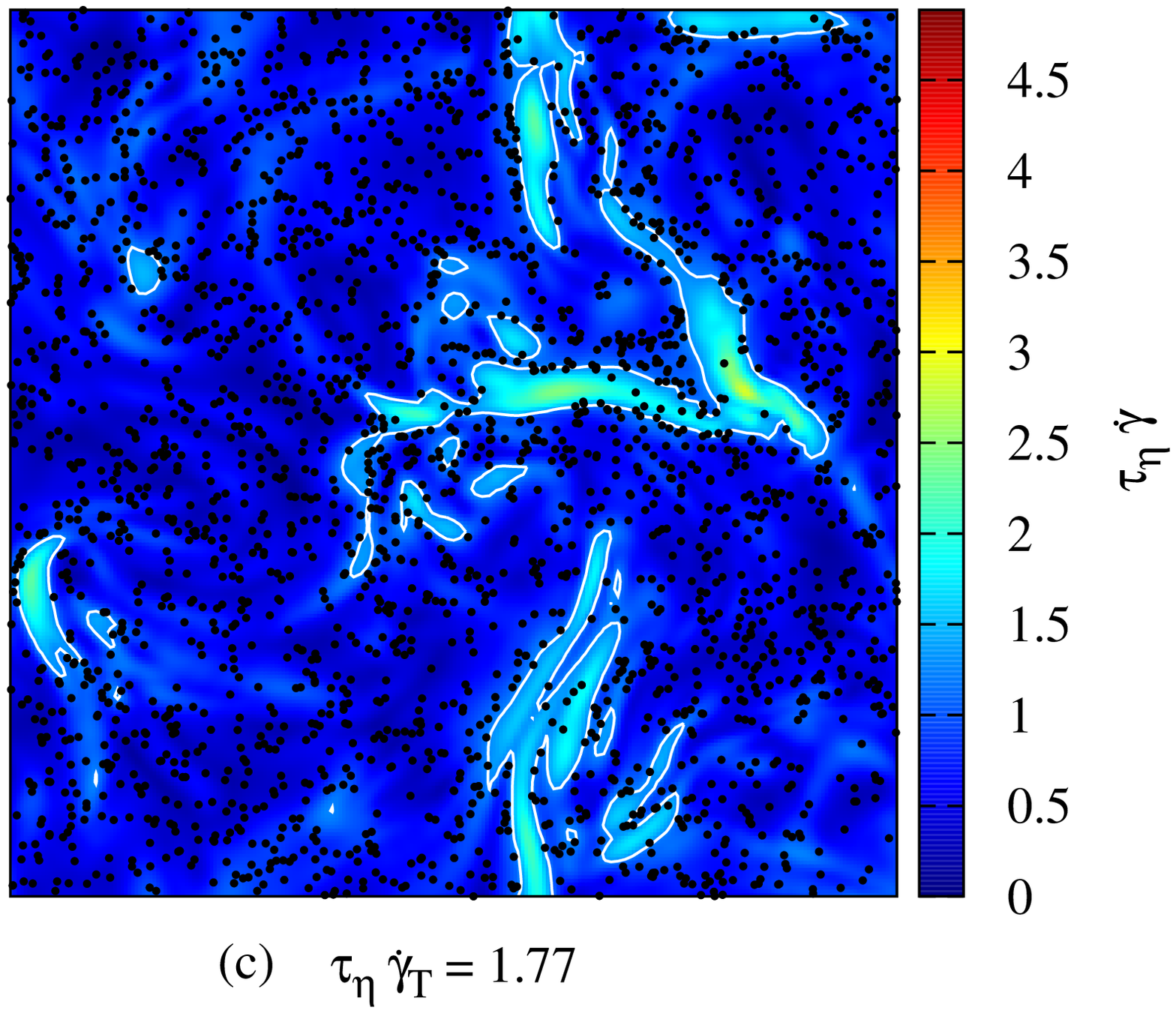}  
    
\caption{Patchiness of the copepods from the simulations. Shading shows the instantaneous field of the absolute value of the shear rate of the Eulerian field, (a) distribution of the copepods with $u_J/u_\eta = 250$ and $\tau_\eta \dot{\gamma}_T = 0.35$, (b) distribution of the copepods with $u_J/u_\eta = 250$ and $\tau_\eta \dot{\gamma}_T = 0.92$, (c) distribution of the copepods with $u_J/u_\eta = 250$ and $\tau_\eta \dot{\gamma}_T = 1.77$. Contour lines are traced at the corresponding value of $\dot{\gamma}_T$ on each panel.}
\label{patchiness}
\end{center}
\end{figure}

The distribution of the LC particles is illustrated by figure \ref{patchiness}, where we show the instantaneous particle positions in two-dimensional slices of thickness $\sim \eta$, visualising at the same time the values of shear-rate of the carrying flow.
Contrary to fluid tracers, LC particles are non-homogeneously dispersed in regions where turbulence intensity is below the given shear-rate threshold, according to the model.  In the panels of figure \ref{patchiness}, we also highlight the $\dot{\gamma}_T$ values by contour lines, we name respectively \textit{comfort} and \textit{alert} regions the locations which are below or above these fixed $\dot{\gamma}_T$ values.  In Fig. \ref{patchiness}(a), which corresponds to $\dot{\gamma}_T = 0.35\ \tau_\eta^{-1}$, the alert region is the dominant one. In this situation the great majority of LC particles are jumping 
but they manifestly fail to reach the few  available comfort islands.  This may be due both to the fact that islands are small and that they are short lived: one shall bear in mind the interplay between space and time in this problem.
The panel (b) shows a condition where comfort and alert regions are equally probable. We notice a pronounced aggregation of particles in the alert areas surrounding the comfort regions, while the latter are efficiently evacuated. 
Finally, the panel (c) illustrates what happens when the alert behaviour is triggered only by few extreme shear rate filamentary regions. The LC particles manage to avoid them quite efficiently but in the overall picture they seems to be mostly homogeneously distributed. (See Supplemental Material \cite{Supplemental-Material} for the 2D visualisation of copepod's motion in turbulent flow at $\tau_\eta \dot{\gamma}_T = 0.92$.)

%\FloatBarrier

\begin{figure}[!htb]
\begin{center}
%%%   trim=top left bottom right 
  \includegraphics[clip, trim=2cm 2cm 1cm 1cm, scale = 0.37, angle=-90]{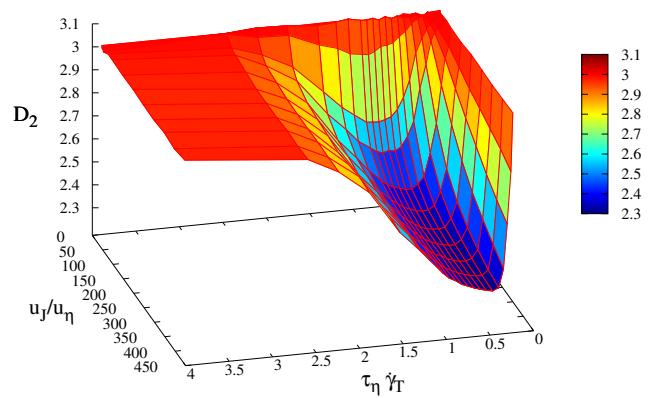}  
\caption{Correlation dimension $D_2$ of copepods as a function of  jump intensity $u_J/u_\eta$ and threshold value $\tau_\eta \dot{\gamma}_T$.}
\label{3D-D2}
\end{center}
\end{figure}

\begin{figure}[!htb]
\begin{center}
  \includegraphics[scale = 0.3, angle=-90]{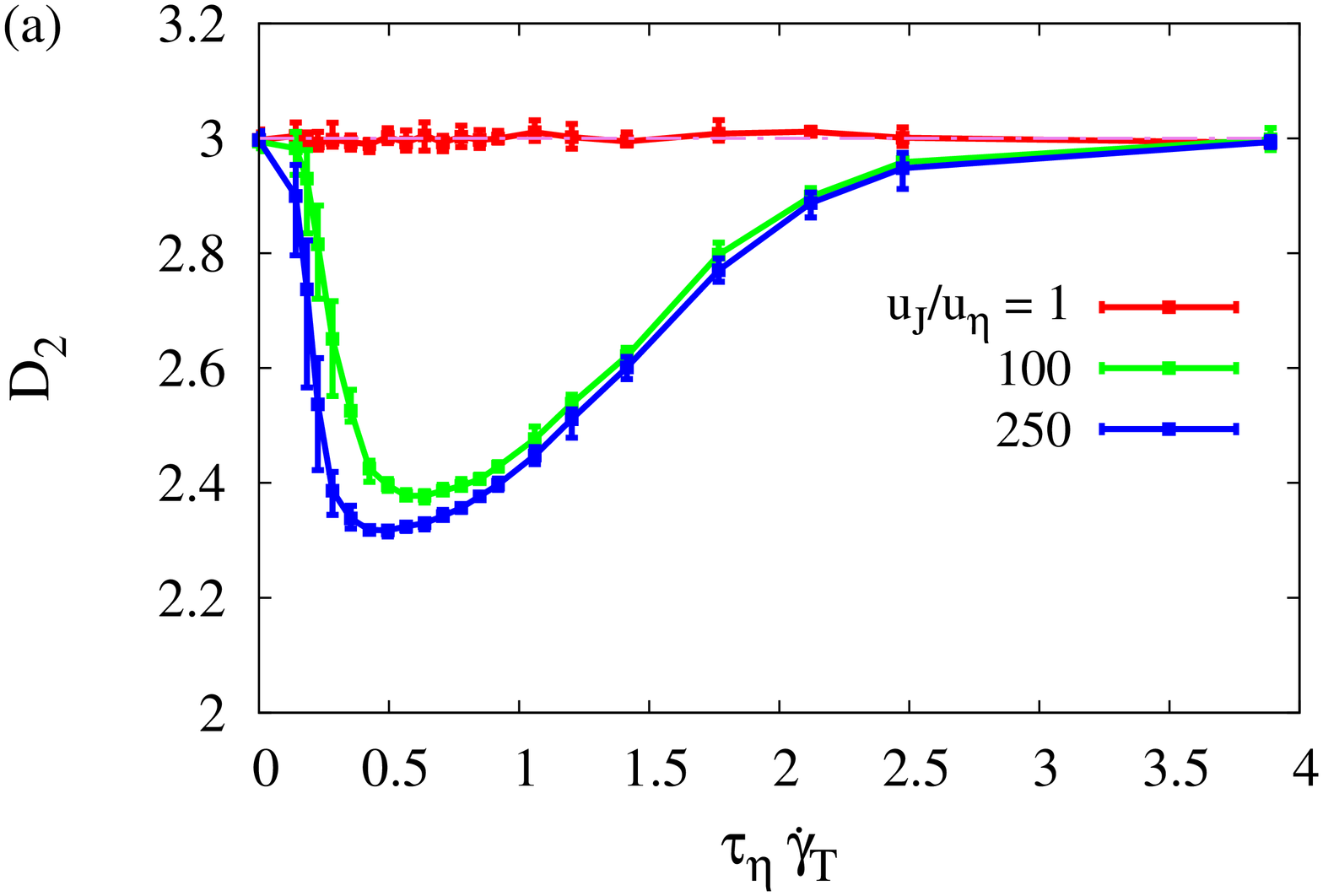}  
  \includegraphics[scale = 0.3, angle=-90]{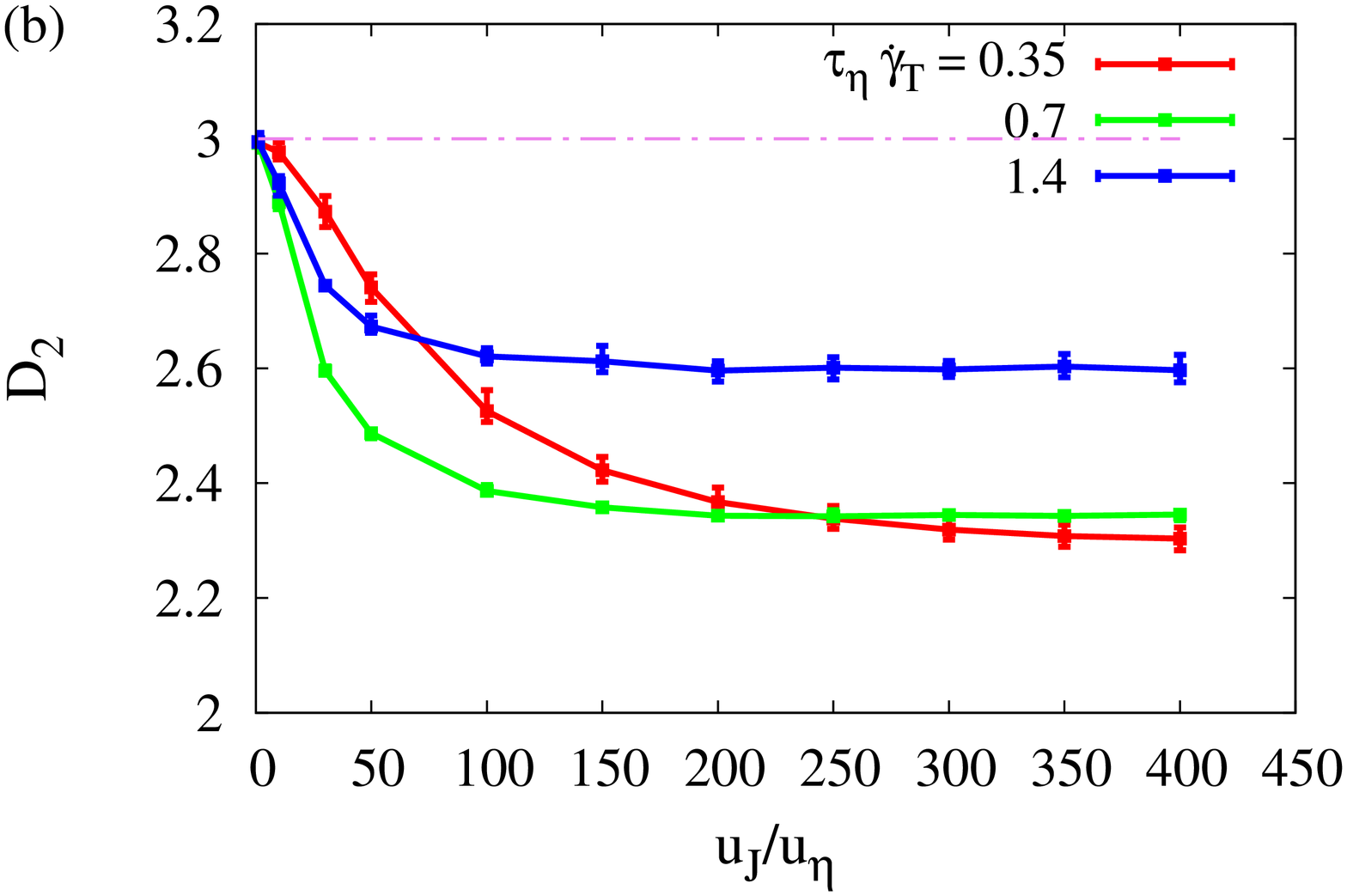}  
\caption{Lateral view of correlation dimension of the copepods as a function of the jump intensity $u_J/u_\eta$ and threshold value $\tau_\eta \dot{\gamma}_T$. Error bars indicate the range of variability of the measurements from 10 independent particle snapshots.}
\label{D2-lateral}
\end{center}
\end{figure}

In order to better quantify the patchiness of the LC particles we compute their Correlation Dimension ($D_2$), which is a measure of the dimensionality of a set of points. According to the Grassberger and Procaccia algorithm \cite{Grassberger-1983}, the $D_2$ is defined as the scaling exponent of the probability of finding a pair of particles with a separation distance less than $r : P_2 (|X_2-X_1|<\,r) \propto r^{D_2}$ as $ r \to 0$. In other words if 
\begin{equation}
C(r) = \frac{2}{N(N-1)}\sum\limits_{i<j} {H(r-|X_i-X_j|)}
\end{equation}
decreases like a power law, then $D_2 = \lim\limits_{r \to 0} \frac{\log C(r)}{\log r}$. Figure \ref{3D-D2} shows the  $D_2$ value in the two dimensional parameter space composed by the intensity of the jump and the shear rate threshold value. The clustering ($D_2 < 3$) is discernible when the prescribed shear rate threshold value is less than $2.8\ \tau_\eta^{-1}$, and it is maximal, $D_2 \simeq 2.3$, at around $0.5\ \tau_\eta^{-1}$. On the other hand we observe a saturation of clustering as $u_J$ is increased. In order to better appreciate these two features, i.e. the minimum with respect to $\tau_\eta \dot{\gamma}_T$ and a saturation as a function of $u_J/u_\eta$, two two-dimensional cuts of the $D_2(\dot{\gamma}_T,u_J)$ surface are shown in Fig. \ref{D2-lateral}.

%\FloatBarrier

One may wonder why there is an optimum and what is its physical meaning. 
Copepods are prone to jump in order to escape from regions of alert $(\dot{\gamma}>\dot{\gamma}_T)$, to reach regions where $\dot{\gamma}<\dot{\gamma}_T$, therefore the chance for a jump to be successful (assuming it to be randomly oriented) depends on the size of the comfort region, in other words to the volume, $\mathcal{V}_{\dot{\gamma}<\dot{\gamma}_T}$. On the other hand, clustering would be maximum if we have numerous successful jumps, and obviously the number of jumps depends on $\mathcal{V}_{\dot{\gamma}>\dot{\gamma}_T}$. This implies that copepods clustering is expected to be proportional to $\mathcal{V}_{\dot{\gamma}<\dot{\gamma}_T} \cdot \mathcal{V}_{\dot{\gamma}>\dot{\gamma}_T}$. Now, substituting the volume of comfortable regions with $\mathcal{V}_{tot} - \mathcal{V}_{\dot{\gamma}>\dot{\gamma}_T}$ leads to $\mathcal{V}_{\dot{\gamma}>\dot{\gamma}_T} \cdot (\mathcal{V}_{tot} - \mathcal{V}_{\dot{\gamma}>\dot{\gamma}_T})$. One direct consequence is that the clustering would be maximum when $\mathcal{V}_{\dot{\gamma}>\dot{\gamma}_T} = \mathcal{V}_{tot}/2$. This can explain the existence of the optimum of $D_2$ as a function of $\tau_\eta \dot{\gamma}_T$ as shown in figure \ref{D2-lateral}(a) as well as its trend as a function of  $\dot{\gamma}_T$. Note however that this argument is based on the simplifying assumption that there is no correlation between the orientation of a LC particle at jump and its position respect to the comfort area, and also it neglects the spatial structure of the shear rate field.

%\FloatBarrier

\begin{figure}[!htb]
\begin{center}
   \includegraphics[scale = 0.33, angle=-90]{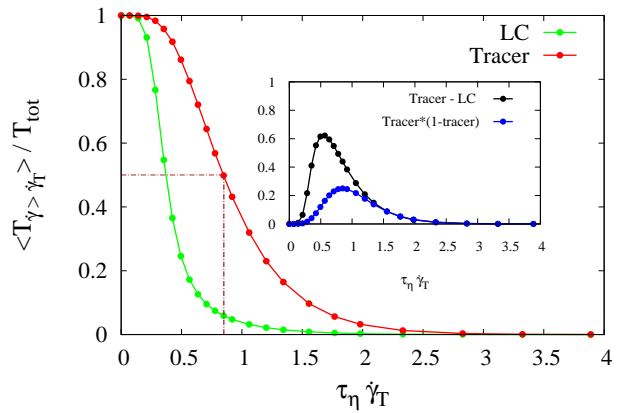}  
\caption{Time fraction spent in alert regions by tracer and copepods as a function of the threshold value. Inset: difference of time fraction between tracers and LC particles, and prediction based on $\mathcal{V}_{\dot{\gamma}>\dot{\gamma}_T}/\mathcal{V}_{tot} $ measurement.}
\label{residence-time}
\end{center}
\end{figure}

How can we determine the value of $\dot{\gamma}$ for which the condition of $\mathcal{V}_{\dot{\gamma}>\dot{\gamma}_T} = \mathcal{V}_{tot}/2$ occurs? One possibility is to perform an Eulerian measurement of the $\dot{\gamma}(x,t)$ field over space and time.
Another option is to look at the fraction of time spent by tracers in alert regions, $T_{\dot{\gamma}>\dot{\gamma_T}} / T_{tot}$ (with $T_{tot}$ the total time of the measurement). Since tracers explore evenly all the region of the flow this is equivalent to measure the volume ratio $\mathcal{V}_{\dot{\gamma}>\dot{\gamma}_T}/\mathcal{V}_{tot} $. 
In particular in order to increase the statistical sampling we look at the global mean value $\langle T_{\dot{\gamma}>\dot{\gamma}_T} \rangle / T_{tot}$ where the average is over the total number of  particles ($N_{tot}$): 
\begin{equation}
 \langle T_{\dot{\gamma}>\dot{\gamma}_T} \rangle  = \frac{1}{N_{tot}}\sum\limits_{i=1}^{N_{tot}}\ \int_{0}^{T_{tot}} H(\dot{\gamma}_i(t)-\dot{\gamma}_T)\ dt.
\end{equation}
The plot in figure \ref{residence-time} shows the trend of $\langle T_{\dot{\gamma}>\dot{\gamma}_T} \rangle / T_{tot}$ as  function of $\dot{\gamma}_T$ both for tracers and LC particles. It confirms that copepods reside less in alert regions compared to tracers. Moreover the difference among the two time fractions can be used as an alternative clustering indicator. It has in fact a similar trend as the $D_2(\dot{\gamma}_T)$ function and shows a peak for the same value of $\dot{\gamma}_T$ (inset of Fig.\ref{residence-time}). The prediction that clustering varies as $\mathcal{V}_{\dot{\gamma}>\dot{\gamma}_T} \cdot (\mathcal{V}_{tot} - \mathcal{V}_{\dot{\gamma}>\dot{\gamma}_T})$ is in qualitative agreement with the observed trend, it is in quite good agreement in the  large $\dot{\gamma}_T$ regime, however it fails to capture the correct value at which the maximum appears, giving $\tau_\eta \dot{\gamma}_T = 0.85$ instead of 0.5. Finally, we note that the case of maximal clustering at $D_2 \simeq 2.3$ corresponds to a condition where the LC particles concentrate in nearly two-dimensional sheets which envelop the \textit{alert} regions (as can be also inferred from the visualisation in Fig. \ref{patchiness}(b)).

%\FloatBarrier

We can offer a qualitative physical explanation for the observed $D_2$ saturation for high values of $u_{J}$ at fixed $\dot{\gamma}_T$, (figure \ref{D2-lateral}(b)).
The argument is as follow: one may expect that there is clustering if the time to escape from an alert region is less than the lifetime of such a region: $\tau_{escape} <  \tau_{\dot{\gamma}_T}$.
The former time can be estimated as $\tau_{escape}=l_{\dot{\gamma}_T}/u_{J}$,  where $l_{\dot{\gamma}_T}$ is the typical size of the alert region characterised by a shear-rate  $\dot{\gamma} > \dot{\gamma}_T$. This implies that LC particles form clusters and the $D_2$ measure is lead to saturate to a constant value if $u_{J} >  l_{\dot{\gamma}_T}/\tau_{\dot{\gamma}_T}$.  This latter ratio can be thought as a threshold dependent escape velocity $u_{\dot{\gamma}_T} = l_{\dot{\gamma}_T}/\tau_{\dot{\gamma}_T}$. From the correlation dimension measurement this escape velocity is estimated to be of the order of  $100 \, u_{\eta}$, \textit{i.e.} of the order of the large scale velocity, with a weak decreasing trend at increasing $\dot{\gamma}_T$.

We finally observe that when the flow field associated to the Lagrangian particles, $\bm{v}=\dot{\bm{x}}$, displays a weak compressibility, it can be shown \cite{Falkovich-2001, Durham-2013} that  $D_2$ depends on the flow divergence by the relation $D_2 = 3 - c \langle \nabla \cdot v \rangle^2$ with $c$ a proportionality constant and angular brackets denoting time and space average. If this argument is applied to the LC model we observe that the divergence can be different form zero only at the interface between comfort and alert regions. This is because in comfort regions ($\nabla \cdot \bm{v} = \nabla \cdot \bm{u} = 0$) and in alert regions ($\nabla \cdot \bm{v} = \nabla \cdot \bm{u} + \nabla \cdot \bm{J} = 0$,  as we can safely assume the jump term to be spatially constant).  At the interface however, the change from the fluid velocity intensity $u$ to $u+u_J$ has a spatial transition scale roughly proportional to $u_J \cdot \log{(10^2)} \tau_J$ which leads to a non-null divergence. This explains the LC accumulation that we observe in correspondence of the alert/comfort interfaces, which effectively acts as sink or source term of the  LC velocity field (see in particular the central panel of Fig. \ref{patchiness}).
By following this line of reasoning, one can guess that the minimum value of $D_2$ will correspond to the case where the surface of alert/comfort interface
is maximum (and not of volumes, as stated above). This has clearly a dependence on the threshold $\dot{\gamma}_T$  and much less, if any, on $u_J$. Despite the qualitative agreement of this observation with our numerical results, we have not been able yet to confirm it quantitively in the weakly compressible limit of the LC model.

\subsection{Particle Orientational Dynamics}
\begin{figure}[!h]
\begin{center}
  \includegraphics[scale = 0.3, angle=-90]{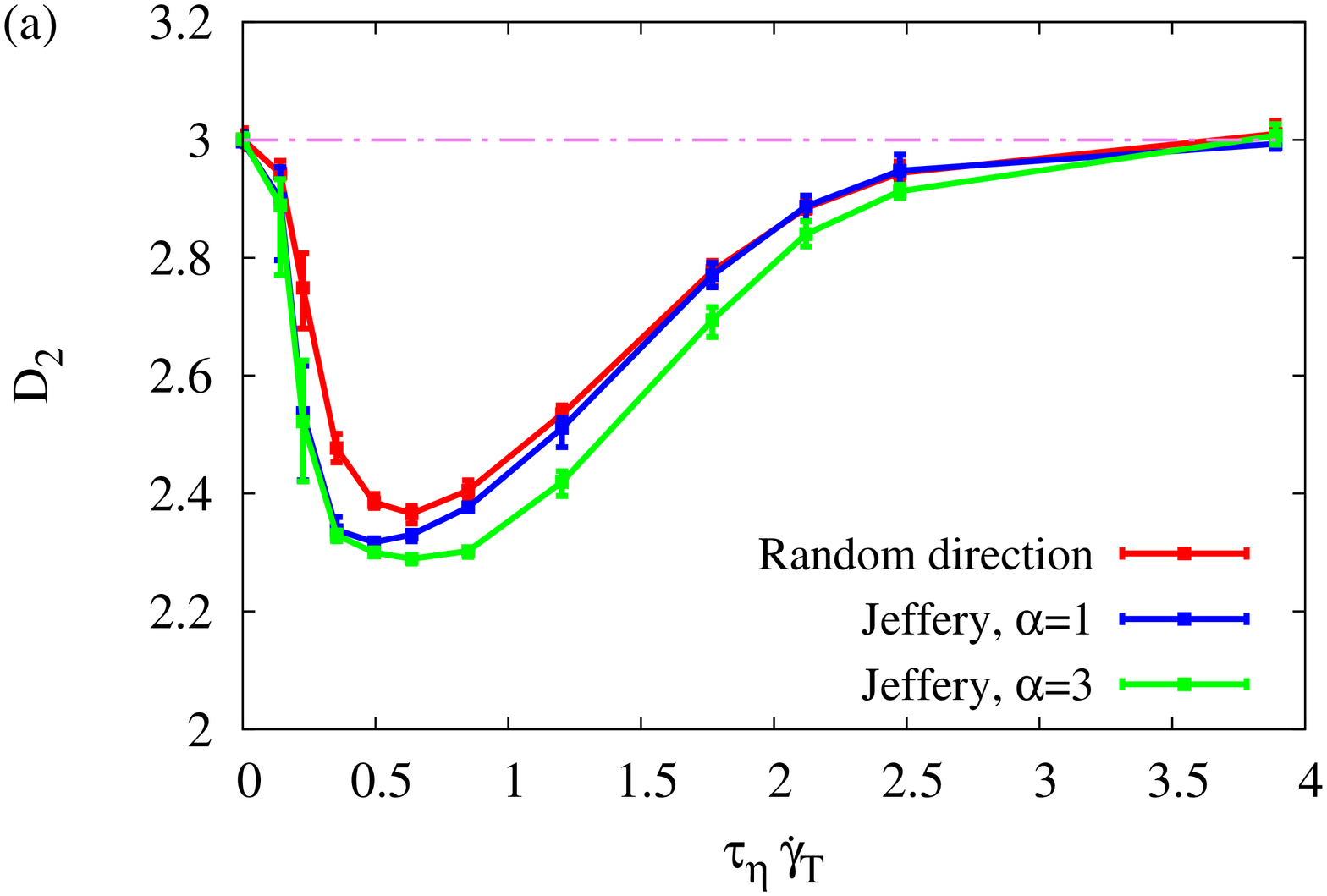}  
  \includegraphics[scale = 0.3, angle=-90]{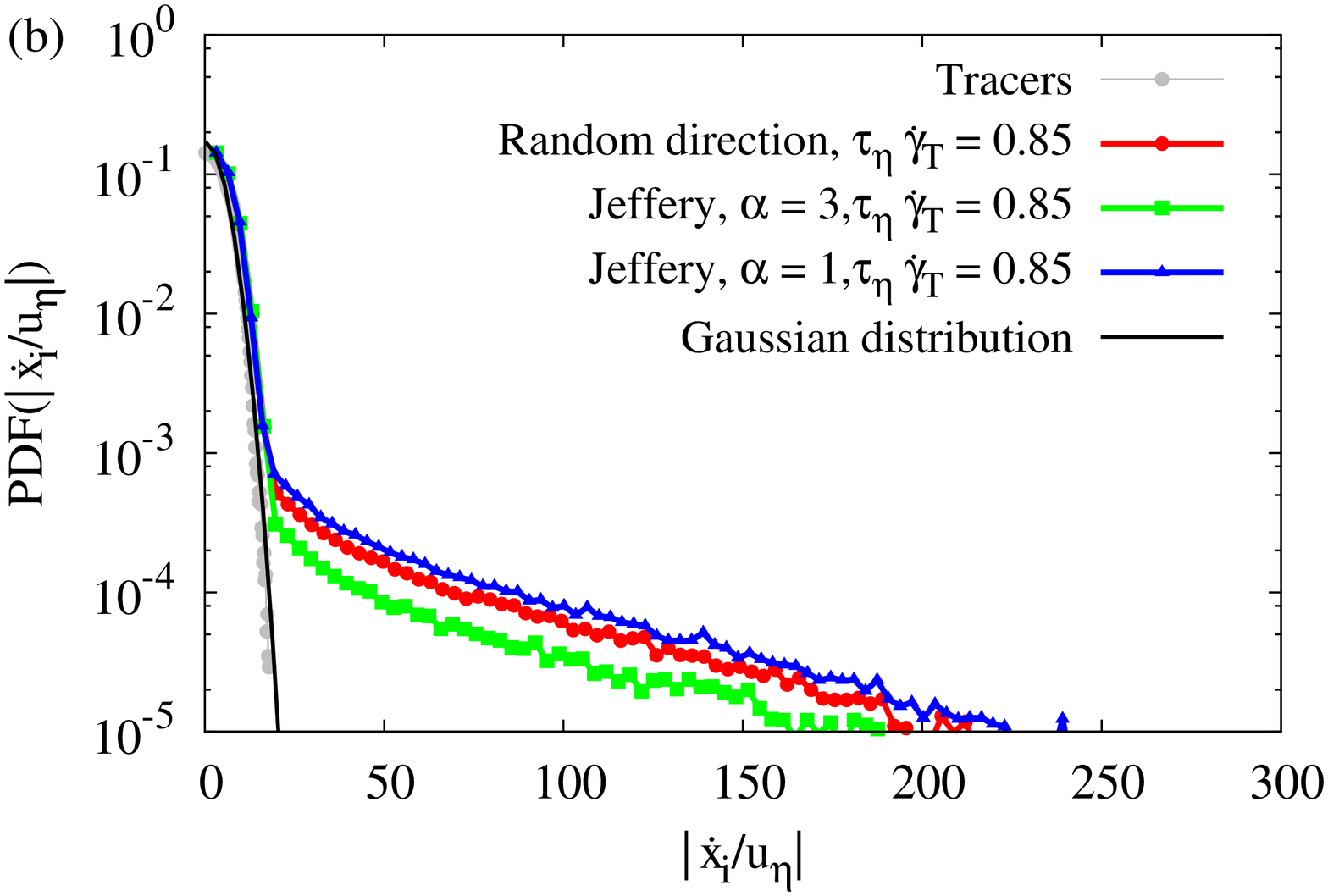}  
\caption{(a) Effect of the aspect ratio and direction of the jump on the fractal dimension, (red) Copepod as solid sphere particles, their direction of the jump is random in the solid angle; (blue) Copepod as solid sphere particles with an orientation and (green) as elongated particles. Both jump in a direction following the Jeffery's equation. (b) PDF of the absolute value of the single component velocity for (red) random direction case, and (green) Jeffery's case with aspect ratio of 3. All cases are computed at $u_J/u_\eta = 250$.}
\label{compare-all}
\end{center}
\end{figure}

What is the importance of particle orientation for the non homogenous distribution of particles? 
The effect of the geometrical aspect ratio of the particles, together with the direction of their jump on the fractal dimension are addressed here. The fluid deformation rate symmetric tensor $S_{ij}$ comes into play by modelling copepods as elongated particles with aspect ratio equal to 3 (e.g. the relevant aspect ratio for \textit{Eurytemora affinis} copepod). Its effect on the jump direction selection leads to enhanced clustering of the particles for jump intensity $u_J/u_\eta = 250$. Copepods can also jump in random direction in the solid angle independently from the rotation rate and deformation rate of the Eulerian field. Less clustering in this case is logical since the jumping direction has no relation with the fluid flow. These behaviours can be found in more details in figure \ref{compare-all}, where we address the influence of jump direction on the PDF of the copepods velocity.

%\FloatBarrier

\section{Conclusions and perspectives}
In this study we have considered a Lagrangian model for active particles. The model is trimmed in a way to reproduce some dynamical features experimentally observed in the motion of copepods in still water. Its main characteristics is the possibility to locally acquire an extra-velocity  (jump)  in response to a variation of the fluid flow conditions surrounding the particle.  The direction of the jump is ruled by the hydrodynamics of small neutrally-buoyant particles. The Lagrangian model has been coupled to a turbulent developed flow described by the incompressible Navier-Stokes equations.\\
We have shown that jump escape reaction from spatio-temporal events characterised by high shear-rate leads to non homogeneous spatial distributions of active particles. This clustering mechanisms however is effective only when the reaction threshold is close to values of the order of $\tau_{\eta}^{-1}$ in a very narrow range. The fact that the range is narrow is ultimately linked to the intermittent distribution of the turbulence dissipation rate \cite{Frisch-book}. We have shown that clustering approaches its maximum when the threshold rate value $\dot{\gamma}_T$ roughly divide the shear-rate $(\dot{\gamma})$ spatial field in equal volume regions. Since this mechanisms mainly depends on the average value of small-turbulence scales rather than on their fluctuations we expect it to have a weak dependence on the Reynolds number of the turbulent flow.  A second implication of the model is that for any given shear-rate reaction value $\dot{\gamma}_T$ there is a maximal intensity jump velocity beyond which clustering can not be further increased.  Finally, the analysis of the correlation dimension suggests the formation of local quasi-bidimensional clusters enclosing the non-permitted flow regions.
From a physicist viewpoint we remark that the clustering mechanism at work in turbulence for the LC model is different form the one shown in other model systems of particulate active matter. For instance the clustering observed for motile algal cells in turbulence is given by the gyrotactic effect, which is a non-isotropic effect induced by the presence of the the external gravity field \cite{DeLillo-2014}. On the opposite, the LC model discussed here is isotropic but it is non-homogeneous in space (it depends on the local value of the shear-rate). We have tested the fact that clustering also appears when LC particles are made sensitive to other flow quantities such as enstrophy or fluid acceleration. The minimal fractal dimension we observed is always above the value of 2, confirming the fact that particles in this case aggregate in order to cover the surface of the forbidden regions. Based on these observations we do not expect that such clustering processes could lead to filamentary like clusters, $D_2 \simeq 1$, as the ones observed for microbubbles in turbulent flows.  Another notable result is the negligible impact of the particle orientational dynamics on the clustering. This is likely to be linked to the limited duration of jumps (note that here $\tau_J \ll \tau_{\eta}$), but might become important for longer jumps, particularly in the modelling of larger motile plankton.  The negligible impact of orientation for the case examined here, suggests the possibility to formulate accurate eulerian mean-field  particle models based on the introduction of a space-dependent effective diffusivity $(\kappa)$ whose amplitude may be linked directly to jump shape parameters,  via a dimensional relation of the type $\kappa \propto u_J^2 \tau_J$.\\
From a more biological perspective, although behavioural mechanisms leading to clustering had been already suggested in the past, such as the formation of patches through swimming against the flow \cite{Genin-2005}, the possibility of cluster formation by escape jumps in a no-mean flow situation was never reported before.
As discussed in \cite{Schmitt-2008}, clustering of copepods has an ecological importance: an effect may be to strongly increase the contact rate with mates, and hence improve the reproduction. Indeed several models have been proposed to express copepod contact rates in turbulence \cite{Rothschild-1998, Evans-1989, Visser-1998}, reviewed in \cite{Lewis-2000}. In case of clustering, the contact rate is strongly increased \cite{Wang-1998, Reade-2000, Collins-2004, Schmitt-2008}. The clustering which would result from a behaviour of predator avoidance (a reaction to turbulent shears similar to predator's signals) would have as side-effect a positive consequence with a strong enhancement of the mating contact rate. 
Of course such copepod concentration could also attract predators. Due to different trade-offs, each copepod species may have an optimal jump behaviour in response to turbulence. For example the copepod \textit{Eurytemora affinis} used in our experimental section is an estuarine species adapted to maintain the bulk of its population in a salinity gradient in highly turbulent conditions \cite{Devreker-2008, Schmitt-2011}. By using high frequency sampling data of all life stages of \textit{E. affinis}, Schmitt et al. \cite{Schmitt-2011} confirmed that the late developmental stages (mainly adults) exhibited active vertical migration during the flood. Consequently the population was not homogeneously distributed in the water column, as dense patches are observed during short time window and near the bottom \cite{Devreker-2008}.\\ 
Our model can be improved in the future to test  such situation with tidally induced turbulence in shallow estuaries where copepods can use their jump abilities to simply avoid to be flushed out their optimal habitat. This could lead to the identification of some optimal clustering strategy that may be in relation with the dome-shapes proposed earlier, on purely speculative intuitions \cite{Cury-1989, MacKenzie-2000}. The presented LC model can also be improved by refining the jumping protocol in order to take into account the fact that the temporal sequence of jumps in copepods occurs in fast sequences (bursts) interposed to inactive moments. Another possible direction of research concerns the investigation of the impact of a spatial radius of perception for copepods to react to turbulent shear. This may produce a smoothing or a delay in the perceived turbulent signal.\\
\textit{Aknowledgments:} The authors acknowledge support from European COST Action MP1305 ``Flowing Matter" and the ``GIP Seine-Aval ZOOGLOBAL project". Dominique Menu is acknowledged for technical help concerning the copepod experiment devices, and Regis Sion for advices and help on the use of the fast camera. Marion Roussin and Gu\'{e}nol\'{e} Alizard are thanked for help with copepod sorting. We acknowledge discussion with Massimo Cencini and Guido Boffetta during the workshop Micoroorganisms in Turbulent Flows, recently held at Lorentz Center in Leiden. H. A. is supported by the PhD grant for interdisciplinary research ``Allocation President 2013" of the University of Lille 1.

%----------------------------------------------------------------------------------------
%	REFERENCE LIST
%----------------------------------------------------------------------------------------
%\bibliographystyle{ksfh_nat}
%\bibliographystyle{apsrev}
%\bibliographystyle{plainnat}
%\bibliographystyle{aipauth4-1}
%\medskip
\bibliography{myref}

\end{document}